# Collimated protons accelerated from an overdense gas jet irradiated by a 1 µm wavelength high-intensity short-pulse laser

S.N. Chen[1,2,3,*], M. Vranic[4], T. Gangolf[1,5], E. Boella[4], P. Antici[6], M. Bailly-Grandvaux[7], P. Loiseau[8], H. Pépin[6], G. Revet[1,2], J. J. Santos[7], A.M. Schroer[5], Mikhail Starodubtsev[2], O. Willi[5], L. O. Silva[4], E. d'Humières[7], J. Fuchs[1,2]

1. LULI, École Polytechnique, CNRS, CEA, UPMC, 91128 Palaiseau, France
2. Institute of Applied Physics, 46 Ulyanov Street, 603950 Nizhny Novgorod, Russia
3. Light Stream Labs LLC., Sunnyvale, CA, USA
4. GoLP/Instituto de Plasmas e Fusão Nuclear, Instituto Superior Técnico, Universidade de Lisboa, 1049-001 Lisboa, Portugal
5. Institut für Laser- und Plasmaphysik, Heinrich-Heine-Universität Düsseldorf, 40225 Düsseldorf, Germany
6. INRS-EMT, 1650, boulevard Lionel-Boulet, J3X 1S2 Varennes (Québec), Canada
7. Univ. Bordeaux, CNRS, CEA, CELIA (Centre Laser Intenses et Applications), UMR 5107, F-33405 Talence, France

**Abstract**

We have investigated proton acceleration in the forward direction from a near-critical density hydrogen gas jet target irradiated by a high intensity ($10^{18}$ W/cm$^2$), short-pulse (5 ps) laser with wavelength of 1.054 µm. We observe the signature of shock acceleration driven by the laser pulse, leading to monoenergetic proton beams with small divergence in addition to the commonly used electron-sheath driven proton acceleration. The proton energies we obtained are modest (~MeV), but prospects for improvement are offered through tailoring the gas jet density profile. Also, we observe that this mechanism is very robust in producing those beams and thus can be considered as a future candidate in laser-driven ion sources driven by the upcoming next generation of multi-PW near-infrared lasers.

*sophia.chen@lightstreamlabs.com

**Introduction**

Over the past decade, laser-accelerated ion beams [1,2,3,4] have attracted considerable interest due to their unique characteristics and have already enabled many applications. These include ultrafast radiography [5,6,7,8] and prompt heating of dense matter [9,10,11]. However other scientific (laser-driven ion fusion) [12], medical (hadron therapy) [13,14,15] or more mainstream (like nuclear fuel recycling through Accelerator-Driven-System) applications can only be unlocked with further progress. Common to all is indeed the need for controllable energy bandwidth, low divergence at the source, and also high repetition rate. The hurdle of a high repetition ion beam can be addressed easily with the increasing repetition rate of presently available [16] and upcoming [17, 18] laser drivers. Lifting the first two hurdles (bandwidth and divergence) is however more difficult as it requires moving away from the presently mostly relied upon acceleration method, i.e. the so-called Target Normal Sheath Acceleration (TNSA) mechanism [19]. This mechanism is very robust, but it intrinsically produces broadband energy (with 100 % spread, unless the number of available ions to accelerate is purposely reduced [20]) beams having angular divergence, variable within the energy spectrum, up to 20° [21, 22].

Several alternative acceleration schemes that would offer the desired characteristics in beam parameters have been already proposed and tested. One of these schemes relies on radiation-pressure driven acceleration (RPA) of ions in ultra-thin targets [23]. It is demanding not only in terms of target thickness and overall stability, but also in terms of laser parameters. Indeed, for RPA, the laser pulse must have ultra-high contrast to not damage the target prior to the main pulse irradiation as well as it must have an ultra-high intensity such that this acceleration mechanism would be dominant with respect to TNSA. For these reasons, with present-day lasers, only the onset of the RPA acceleration mechanism, mixed with TNSA, could be demonstrated [24,25,26] and questions, relative to the beam quality being possibly affected by target instabilities [27], still remain. Another promising scheme was introduced by Denavit et al. [28], followed by Silva et al. [29] for critically dense targets known as Collisional Shock Acceleration (CSA). It is based on the fact that the laser pulse can induce a collisionless shock wave in a near-critical density target, and the propagating shock can reflect incoming ions in the target accelerate them to high energies. Such collisionless shock wave is generated following the injection in the target, beyond the critical density interface at which the laser is stopped, of laser-accelerated fast electrons. Due to their high energy, the collisional dissipation onto these electrons is negligible [30], however collisionless (i.e. mediated by instabilities and plasma waves) processes can provide enough energy dissipation [31]. Thus, a density steepening can form as the fast electrons overcome the target medium in their propagation [29]. In a partially expanded target having near-critical density [32, 33], we note that a variant of TNSA can also take place. As the laser can propagate fully through such expanded target, fast electron currents generated near the target rear surface form a long-living quasistatic magnetic field there. This field generates an inductive electric field at the rear plasma-vacuum interface that complements TNSA in providing ion acceleration [34,35,36] in this so-called Magnetic Vortex Acceleration mechanism (MVA). Finally, note that the laser radiation pressure also directly generates a density pile-up at the critical density interface at which the laser is stopped, the so-called hole-boring (HB) [37], which can also reflect and accelerate ions. We will however show below that, in the conditions explored here, the ions accelerated by CSA have higher energies than that accelerated by HB as the inward shock has a higher velocity than the interface at which the laser is reflected.

Several numerical studies have been performed to optimize the target and laser parameters for CSA, and have shown that targets having peak densities close to the critical density with smooth gradients [39, 38] represent optimal conditions. CSA was then extended to under-critical density targets by d'Humières et al. [39]. There, the shock wave is not created by the laser, but ions-driven in a downward density gradient. This low density CSA (LDCSA) scheme was demonstrated experimentally [40,41] to produce low divergence, yet broadband beams since sheath acceleration in the rear end of the target density profile procures additional acceleration.

Compared to TNSA or RPA, there are other significant advantages to laser-driven CSA other than the prospects of low divergence and monoenergetic beams. First, the scaling with the laser energy of CSA is more favorable than that of TNSA, namely, the ion energy scales linearly with laser intensity [42], whereas for TNSA it scales with the square root of the laser intensity [43]. The second point is purely practical since with TNSA or RPA, solid targets are used and require precise target alignment for each shot, need strict laser temporal contrast, and produce debris in the target chamber. With CSA, especially using gas jets as targets, operation would significantly be easier at upcoming high-repetition rate laser facilities, eliminating the need for target replacement and realignment. Moreover, using a lower-than-solid density for the target would reduce the amount of debris generated [44]. We note that continuous operation of gas jets in high-vacuum chambers have been shown to be possible [45], hence eliminating this concern.

This collisionless shock acceleration (CSA) scheme has been clearly demonstrated experimentally using $CO_2$ lasers [46]. Indeed, the long wavelength (10.6 µm) of these lasers allows for controlled near-critical targets to be easily produced. As mentioned, the laser-driven CSA mechanism is most efficient in a critically dense plasma where $n_e \geq \gamma n_{cr}$ [47], with $n_{cr} = \varepsilon_o m_e \omega_{laser}^2/e^2$ where $\omega_{laser}$ is the angular frequency of the laser, and $\gamma = \sqrt{1 + a_0^2}$ is the relativistic factor for the electrons derived from one-dimensional energy and momentum flux conservation, with $a_0 = (I_L \lambda_L^2/1.37 \times 10^{18} \text{ W}.\mu m^2.\text{cm}^{-2})^{1/2}$ the normalized laser field, $I_L$ and $\lambda_L$ being, respectively, the laser intensity and wavelength. In practical units, $n_{cr}[\text{cm}^{-3}] = 1.1 \times 10^{21}/\lambda_L^2[\mu m]$. Since the wavelength of $CO_2$ lasers is 10.6 µm, the minimum required target density to be overcritical in these conditions is $1 \times 10^{19}$ cm$^{-3}$, which is easily created with commercially available gas bottles and a pulsed valve [48]. Using these targets, it was shown that CSA could indeed generate monoenergetic proton beams, i.e. having less than 5% energy bandwidth, of low (<100 mrad) divergence. The major downside is that in practice $CO_2$ lasers are limited to irradiances $I_L \lambda_L^2$ around $10^{19}$ W.µm$^2$/cm$^2$.

Near-infrared (0.8-1 µm wavelength) lasers exist already at larger irradiances when compared to $CO_2$ lasers, with $I_L \lambda_L^2$ reaching already more than $10^{21}$ W.µm$^2$/cm$^2$ in several facilities world-wide, with prospects for currently built facilities to reach $I > 10^{23}$ W/cm$^2$. However, the difficulty there with respect to CSA is that higher density targets are required, i.e. with densities above $10^{21}$

cm$^{-3}$ ($n_{cr}$ for a 1 μm wavelength laser). This is already possible to achieve with foams [49]; it becomes nowadays possible with gas jets [50,51].

In this article we will show that using a Hydrogen gas jet with a peak density of *2.7$n_{cr}$*, which is irradiated by an intense, short-pulse laser having a wavelength of 1.054 μm, proton beams that display the characteristics of CSA-accelerated beams are observed. With good consistency, the beam displays a mono-energetic peak up to 1 MeV having a very low divergence. The energy of the peak is observed to correlate well with the laser intensity and the target density. The outline of the article is as follows. We will first describe our experimental setup and the measured results. Next, we will present numerical simulations that bring insight into the interaction conditions, notably suggesting that the target width was affected by the prepulse accompanying the intense laser pulse. We will also present results of numerical simulations of the interaction, which reveal that HB and CSA are both at play, but in which CSA is shown, in the conditions of the experiment, to produce higher energy ions (here protons) than HB acceleration mechanisms at play. Moreover, the energy of the ions generated by CSA are found to be in reasonable agreement with the measured ones. Finally, we will discuss prospects for future improvement of such acceleration technique, still with near-infrared lasers, using tailored jets. We note that the recent result of Helle et al. [52] exploits as well a high density hydrogen gas jet and a near-infrared (800 nm wavelength) laser for directed proton acceleration. There, the density is increased by the generation of hydrodynamic shocks induced by auxiliary laser beams and the acceleration is induced by a magnetic vortex. This differs from our results which, when obtained at higher plasma densities, i.e. > 2 $n_{cr}$, are rather related to CSA, as suggested by our simulations.

**Experimental Setup**

The experiment was performed using the Titan laser at the Jupiter Laser Facility (LLNL, USA), using the experimental setup shown in Fig. 1. The short pulse laser arm of Titan, focused with an F/3 off-axis parabola, irradiated a high density gas jet with a maximum of 210 J in 5 ps at a wavelength of 1.054 μm. The laser had a best focus of 10 μm (at its full width at half maximum, or FWHM), containing an encircled energy of around 35%, thus giving a peak intensity at best focus of $I_L \lambda_L^2 = 2.2 \times 10^{19}$ W.μm$^2$/cm$^2$, i.e. yielding the parameter a$_0$=4.2. With these parameters, $\gamma n_{cr}$ =4.3×10$^{21}$ 1/cm$^3$, which a priori sets a very high density requirement to efficiently drive CSA.

Proton spectrometers were placed, as indicated in Fig. 1, on the horizontal plane to measure the proton beam energy and angular distribution. Since we use a pure Hydrogen gas (H$_2$), the spectrometers were not equipped in a Thomson parabola configuration, i.e. they use a simple magnetic deflection to resolve the proton energies. This allows also to use at the spectrometer entrance a wide slit (horizontally) to resolve, for each spectrometer, the proton beam over ±100 mrad around its mean angle of observation. In Fig.1 is shown only the spectrometer located at 0° with respect to the laser incident axis. Other spectrometers were located at 21°, 43°, and 92° with respect to the same axis. As detectors, we use absolutely calibrated [53] FujiFilm image plates of type TR.

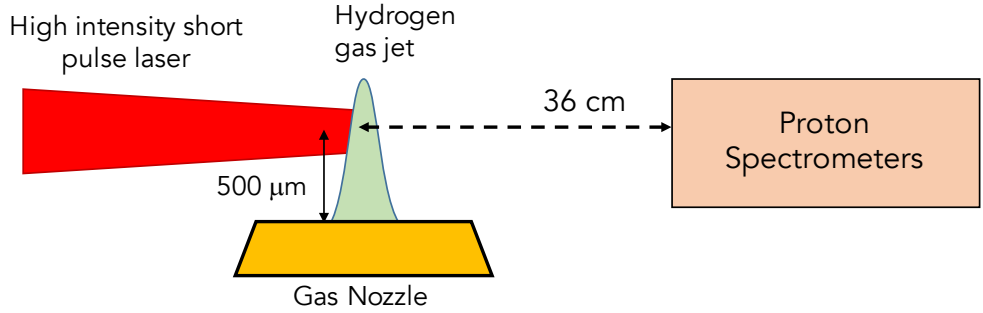

*Figure 1: Experimental Setup.*

The nozzle that we used for the gas jet is a Laval type design to achieve supersonic gas outlet velocity [48]. The orifice was rectangular: 1 mm wide and 300 μm long with a throat of 300 x 300 μm$^2$ located 3 mm below the opening. Before the experiment, we performed 3D optical (using a He-Ne, 633 nm wavelength laser probe) tomography measurements to characterize the rectangular gas profile in the output of the nozzle using Argon gas. It should be noted that the difference in gas flow found by our measurement and others [54] between Argon, a monoatomic gas that was used in the test, and Hydrogen (H$_2$), a diatomic gas that was used for the experiment has a difference of profile and molecular density of less than 1%. This is consistent with calculations that can be made of the gas flow in the exit of the nozzle [55] and which suggest that the differences between Ar and H$_2$ flows (having respective specific heat ratio 7/5 for for H$_2$, and 5/3 for Ar) are quite minor. Fig. 2 shows a horizontal cross-section of the gas density distribution at 500 μm from the base of the nozzle. We measured, in the range of 10 to 100 bars, that the backing pressure is linearly related to the peak density of the gas jet, as demonstrated experimentally [48,56,57]. Measurements at higher pressure are difficult because the high-density in the jet induces first refraction of the optical probe and even, for very high pressures, fully prevents the probe to penetrate the gas jet.

As shown schematically in Fig. 1, the laser was focused at the rising edge of the Hydrogen gas jet and along the narrow part of the density profile (see Fig. 2a). Since the density profile is Gaussian, we chose a position in this profile for the location of the best focus of the laser, which was placed at 150 μm in front of the location of the peak density. The height of the laser focus was 500 μm above the gas jet nozzle, i.e. corresponding to the density profiles shown in Fig. 2b.

To produce the high backing pressures needed to obtain near-critical densities in the gas jet, and starting from a commercially available gas canister (pressurized at 100 bars), we used a Haskel pneumatic gas compressor [58] able to compress the gas to 1000 bars, the Clark Cooper EX40 electro-valve [59] that is rated for these high pressures, and high pressure pipes, fitting and feedthrough from Swagelok [60]. The gas that we used was Hydrogen (H$_2$), i.e. a diatomic molecule at room temperature. Thus, once ionized by the laser, the peak ion density during the interaction is double the molecular density that is shown in the measurements of Figure 2. Thus, by extrapolating our measurements to backing pressures between 150 bars to 900 bars of gas, we conclude that we can *a priori* vary the ionized electron density up to 2.7$n_{cr}$.

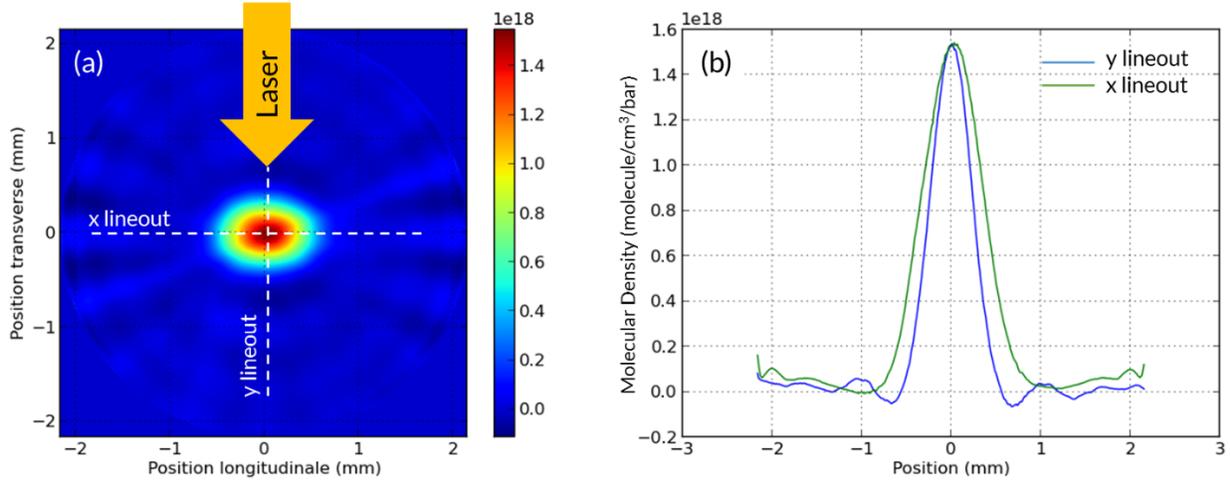

*Figure 2: 3-D tomography of the gas jet. (a) The horizontal cross-section of the gas jet at 500 µm above the nozzle. The colorscale units are in molecular density (molecule/cm$^3$/bar) (b)Vertical and horizontal lineouts of the image in (a), which show that the form of the gas jet can be represented as a quasi-perfect Gaussian.*

**Experimental Results**

Before presenting the proton acceleration results, we should note that the laser pulse that we used to drive the ion acceleration has a pedestal before the main pulse arrives. This pedestal, or prepulse, as measured during the experiment with fast diodes and a water-switch cell, contains around 20 mJ of energy (at the target chamber center (TCC), i.e. at the location of the short-pulse focus) and is characterized by a short ramp (~0.3 ns) preceding a ~1 ns flat plateau, which itself precedes the main pulse. The calibration of the measurement was made by sending a low-energy, 3 ns duration pulse through the chain and the compressor, and measuring its energy simultaneously at TCC, and on the diode which measures the prepulse on every shot. Note that these prepulse values are similar to the ones measured in other runs at the same laser facility by other groups [61,62]. Since the prepulse intensity ($I_L \lambda_L^2 = 10^{13}$ W.µm$^2$/cm$^2$) is above the ionization threshold, it modified significantly the gas jet density profile ahead of the main pulse irradiation. This was on one hand beneficial, since it reduced the thickness of the gas target, which increases the efficiency for CSA, but on the other hand, it had the detrimental effect to push the critical density interface away, i.e. to effectively defocus the high-intensity laser pulse arriving on that interface and reduce its ability to drive a strong shock.

The modification of the gas target profile induced by the laser prepulse is determined by hydrodynamic simulations of the gas jet evolution when it is irradiated by the prepulse. Here we rely on hydrodynamic simulations, using the well characterized ns-duration, low-intensity prepulse of the short-pulse, to infer the target density profiles at the time of the short-pulse irradiation. Indeed, we could not optically probe the interaction due to the over dense gas jet and would have needed an x-ray source (or a second short pulse to create an x-ray burst) to radiograph the plasma. Nevertheless, hydrodynamic simulations in these conditions are well-benchmarked and are able to grasp quantitatively the plasma evolution; such procedure of relying on hydrodynamic simulations has indeed been validated quantitatively many times over the years, as shown e.g. in Refs [40,63,64,65,66]. For such simulations, we used the FCI2 hydrodynamic code [67] in 2D, modelling the same *xy* plane as shown in Fig. 2a. Fitting the measurements shown in Fig. 2, the profile of the gas jet used in the simulation was a Gaussian, as mentioned in the previous section, with a FWHM of 400 µm and using fully ionized Hydrogen with a temperature of 300 K. In the hydrodynamic simulations, the laser propagation from the near field (the focusing optics) to the far-field (focus) and after, is described by a 3D ray-tracing package [68]. We specify a power law that fits the experimental laser power of the pre-pulse in order to get the right laser energy. At each time step, the power is distributed over the rays, then each ray propagates inside the plasma and deposits its power via inverse Bremsstrahlung. Then, a nonlocal electron transport model is used for modeling heat fluxes. The focal plane is adequately defined in terms of spatial dimensions, but ray-tracing packages (based on geometrical optics) do not take into account diffraction. This modelling [69] is sufficient in many situations for describing plasma heating and is widely used in radiative-hydrodynamics ICF codes that have been well benchmarked [70,71]. In our simulations, the box of which is 1.2-mm long and 400-microns wide, we set an initial density profile that fits the gas-jet's longitudinal profile (as derived from Fig.2 of the paper) and we set initial temperatures to an arbitrary low temperature. FCI2 being a radiative-hydrodynamics code designed for describing plasma heating and dynamics, the lower temperature bound used for calculating ionization is around 1 eV, leading to a fully ionized plasma in the whole simulation box, even far from the focal spot. But, this has no incidence on the fact that plasma heating is localized and on the formation of a blast wave: the plasma is still cold far from the focal spot.

The results of the prepulse irradiation of the gas jet are shown in Fig. 3 at various times after the prepulse had begun. We observe that it significantly modifies the gas jet profile, reducing it to about half its initial thickness after 1 ns. As a consequence, the main laser pulse will encounter the remaining steep and dense gas jet interface while being defocused by ~150 µm. Since the focusing optics of the laser is F/3, such defocus results in a reduced intensity at this location of around $3\times10^{18}$ W/cm$^2$, $a_0$ ~ 1. This is estimated by analysing a set of images of the short-pulse beam, as focused by the F/3 parabola, taken at various positions around the best focus. The defocusing is seen to follow very well the theoretical estimate for a Gaussian beam, and we observe that a defocus of 150 µm corresponds to a nominal increase of the beam FWHM from 10 µm (at best focus) to ~45 µm. Apart from such peak laser intensity condition, we also varied (reduced) during the experiment the laser energy or moved the laser focusing point further to the foot of the gas density profile, hence further reducing the laser intensity on the critical density interface. These various conditions will be summarized below.

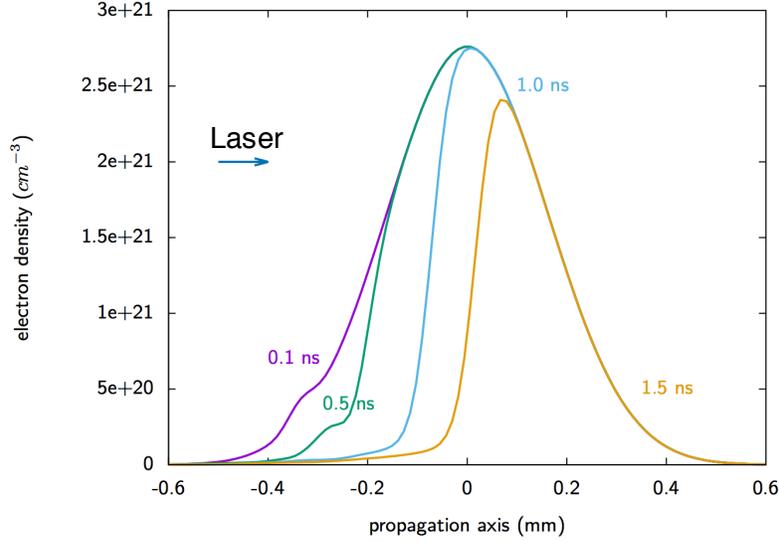

*Figure 3: Hydrodynamic simulation of the spatial profile of the ion density of the Hydrogen gas jet at various times (as indicated) after the start of its irradiation by the prepulse of the Titan laser pulse used during the experiment.*

Experimentally, we first performed a series of shots using the setup shown in Fig. 1. While keeping the laser intensity constant, we varied the backing pressure in the gas jet up to 900 bars, which is the equivalent to varying the peak electron density in the ionized target gas jet from $0.5 n_{cr}$ to $2.7 n_{cr}$. The resulting proton spectra measured with the spectrometer oriented at 0° are shown in Fig. 4. Note that the 100 keV lower end of the spectrum is the instrumental lower detection limit.

As seen in Fig. 4, as the density of the gas target is varied from underdense to overdense, the proton spectrum clearly shows that the energy of the peak in the spectrum increases with the target gas density. Also shown in Fig. 4 are the angular patterns of these proton beams, all displaying a narrow distribution and well resolved within the acceptance of the single spectrometer located at 0°. We stress that in all cases, no signal was recorded in the other spectrometers located at larger angles around the chamber (i.e. there was no signal above the noise level). In the case of overcritical densities, due to the simultaneous observations of a peak in the spectrum, and of a narrow angular distribution for the accelerated beam, we conjecture that the dominant acceleration mechanism could be CSA, as in the case of the $CO_2$ laser experiments. As will be detailed below, we find that this scenario is supported by the numerical simulations.

A clear spectral peak cannot be distinguished in the case of the peak density of $0.5 n_{cr}$, although the angular pattern of the beam in this case displays a narrow profile, even narrower than for higher gas densities: the divergence of this beam is 13 mrad. This extremely small divergence could be due to the MVA mechanism discussed above, i.e. to a quasi-static magnetic field on the back side of the target formed by the hot electrons accelerated directly by the laser on the front side and by the resulting return current [30]. We note that experimentally, proton beams with small divergence have also been observed before by Willingale et al. [72] from underdense gas targets accelerated by the TNSA/MVA process.

As the density of the target is increased to 1.4$n_{cr}$, the spectrum however clearly changes with a significant peak in the proton beam spectrum appearing at 0.4 MeV (with ΔE/E ~0.3). This appears to be a combination of acceleration mechanisms where there is a quasi-monoenergetic beam on top of what appears to be TNSA accelerated protons at lower energy. When the density of the gas jet target is further increased to 2.5$n_{cr}$, the peak has shifted to several hundred keV higher in energy (with ΔE/E ~0.16).

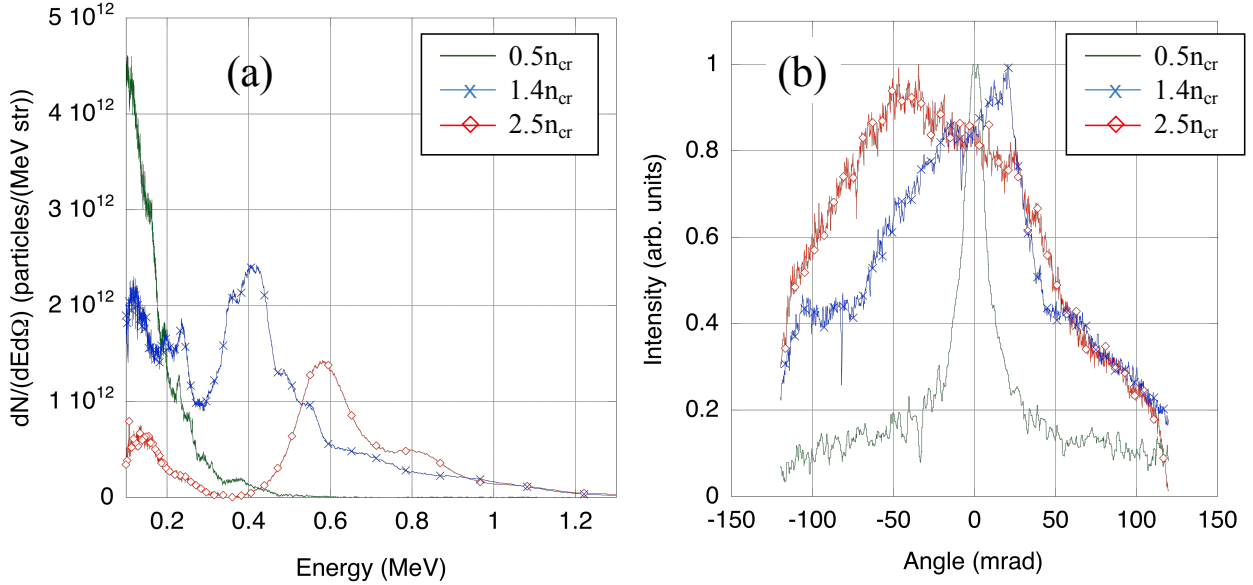

*Figure 4: A comparison of experimentally measured proton beams generated from the gas jet target in the forward (0°) direction and with different peak density for the gas jet target. (a) Proton spectra measured at 0°. (b) Divergence of the proton beam as measured continuously along the slit of the spectrometer located at 0°, and at the location of the proton spectral peak that can be observed in (a). Note that outside the central axis (0°), the spectrum has also a similar shape as shown in (a), with a spectral peak around the same value as at 0°, but with much lower proton number.*

The energy that the protons can acquire through CSA, HB, and TNSA can be estimated using analytical expressions presented by Wilks et al. [37], Fiuza et al. [42] and Stockem-Novo et al. [73]. For these theoretical studies, the final accelerated proton energy, i.e. acquired as the ions are reflected off the shock [29] or hole-boring potential, can be expressed in terms of $\sqrt{I/n_i}$. For the energy of the ions accelerated by the HB, we use $m_i(v_{hb})^2/2$, where $m_i$ is the ion mass and $v_{hb}$ is the HB ion velocity as expressed in Ref. [37], i.e. $v_{hb} = c\sqrt{m_e n_{cr} I\lambda^2 / 2 m_i n_i 1.37 \times 10^{18}}$, where λ is in microns and I in W/cm$^2$. Using a number of shots recorded during the experiment with various laser intensities and gas jet densities, the energy of the quasi-monoenergetic proton beam is plotted against the experimental inputs in the expression $\sqrt{I/n_i}$ in Fig. 5a. The parameter space that we could explore during the experiment was limited due to the low number of shots allocated for a campaign. This low shot number and variability in the laser parameters affects our ability to demonstrate reproducibility. Nonetheless, we can state that the robustness of the process generating peaked spectral distribution at high densities is, with the limited shots we could get on Titan, good, as witnessed by the spectra shown in Fig.4 and Fig.5.

The first curve of green dots represents the expression presented by Fiuza et al. [42] where this expression is dependent on the velocity of the electrostatic shock created by the hot electrons. The shock velocity is there $v_{sh,F} = (2Mc_s)/(1+M^2c_s^2/c^2)$ where $c_s$ is the upstream sound speed, $c$ is the speed of light and $M \sim \sqrt{(1+\eta)/0.4}(n_{cr}/n_i)^{1/2}a_0^{1/2}$ [29] is the shock Mach number. For our experimental intensities, we took $\eta$ to be 0.2 [74], which is the absorption efficiency at the critical density surface at the (defocused) laser intensity we use. The blue curve in Figure 5a represents the expression presented in Stockem-Novo et al. where their expression is based on the velocity of the adiabatic expansion of a gas; this model looks at a shock driven 3D spherical expansion, i.e. it should lead to an underestimate of what we obtain since we work more in a condition closer to a planar shock driven in the gas jet by the high-intensity laser. Here, the velocity is $v_{sh,SN} = c\sqrt{Zm_e n_{cr} a_0^2/8m_i n_0}(1+K_{ad})$ where $K_{ad}$ = 7/3 for diatomic molecules.

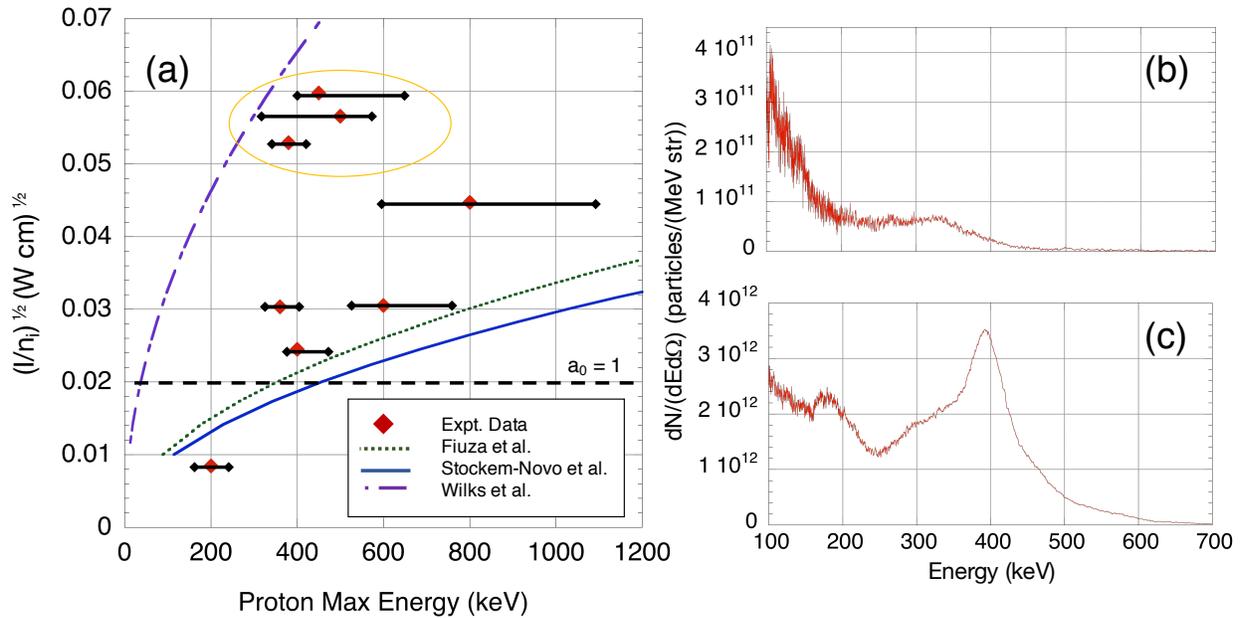

*Figure 5: (a) The proton energy of the quasi-monoenergetic peak recorded in the spectrum for various shots (red dots, as measured in the experiment) for different values of the parameter $(I/n_i)^{1/2}$ and as compared to the hole-boring model (Wilks et al. [37]) and two CSA models (Fiuza et al. [42] & Stockem-Novo et al. [73]). The red center point indicates the energy of the largest population (spectral peak) of protons. The horizontal error bar corresponds to the FWHM of the signal around the spectral peak that is observed. (b) A typical spectrum corresponding to the points encircled in yellow in (a), which correspond to shots recorded at lower pressures, and which are observed to fall near the curve for the hole-boring model. (c) A typical spectrum of the other data points, recorded at higher pressures, which are closer to the CSA scalings.*

As shown in Figure 5a, we can observe that the experimental data fall either close to the curve corresponding to HB or to the one corresponding to CSA. We indeed observe that the data points encircled in the yellow line, close to the HB scaling, have a typical spectrum shown in Figure 5(b) where, on top of an exponentially falling spectrum, there are one or more small peaks on a plateau. These data points correspond to shots at a lower density of the gas jet (i.e. close to $n_{cr}$), hence they are higher in the curve because the gas density ($n_i$) is lower. Contrasting this, the other experimental points with a typical spectrum with a strongly pronounced spectral peak as found in Figure 4a and also shown in Figure 5c, follow the curves describing CSA. These points have been obtained at higher gas densities (i.e. >2 $n_{cr}$), hence they correspond to lower positions in the graph as the factor $(I/n_i)^{1/2}$ is lower. In short, the protons accelerated at high densities, and which display a strong spectral peak, have higher energy than what is predicted by the hole-boring acceleration mechanism, and are close to the CSA trend.

**Numerical Simulations**

To verify that CSA is indeed the proton acceleration mechanism inducing the strong spectral peaks observed in our experimental conditions at high densities (see Fig.4a and 5c) and to gain insight into the actual acceleration processes, we performed particle-in-cell (PIC) simulations using the code OSIRIS [75] in 2D. The simulation box is 1273 μm long and 16 μm wide, resolved with 48000x600 cells and 8 particles per cell. The total simulation time is 20 ps, sampled with a time step of *Δt=0.06 fs*. The initial plasma profile, with peak density of *2.7$n_{cr}$*, used for the first set of simulations is shown (in blue) in Fig. 6. It corresponds to the modified gas jet target profile as found by the hydrodynamic simulations shown in Fig. 3 for the 1 ns duration irradiation of the gas jet by the prepulse. The simulation box is transversely periodic, and the laser is launched from the left-hand wall. The laser is transversely a plane wave, with a temporal envelope of 5 ps at FWHM. The maximum laser intensity reaches the center of the gas jet *(x=0)* at *t=6.5 ps* from the beginning of the simulation.

In general, we note that due to the quasi-1D geometry employed in the simulations, we can expect that the proton energies will be overestimated, especially in the case of TNSA protons [56]. We underline that multi-dimensional simulations of this setup are beyond the current computational capabilities. However, even if one would be able to perform a full-scale 3D simulation of the interaction, it would not be possible to guarantee the quantitative agreement in the proton energies between the PIC simulations and experiment, because this result is sensitive to the differences in the thickness of the initial plasma profile. An additional difficulty rises from the fact that the peak plasma density of the gas jet is close to the relativistic critical density, and small variations in the laser intensity might change the longitudinal position of the critical density interface. Nonetheless, as will be detailed below, the picture described by the simulations is found in reasonable agreement with the experimental data, hence allowing us to believe that the physics observed in the simulations, highlighting shock acceleration as the main mechanism at play, is adequate.

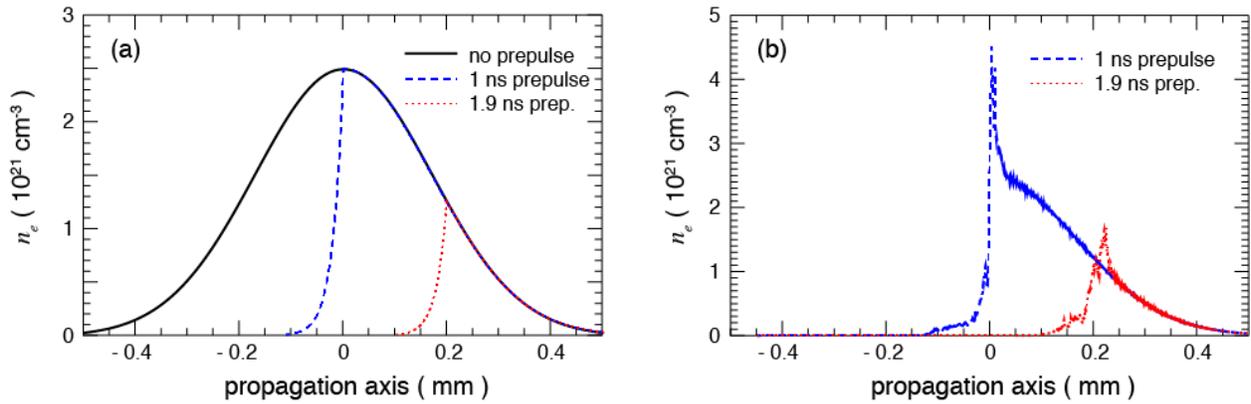

*Figure 6: (a) The density profiles are calculated by FCI2 and used as the input into the PIC simulation. Longitudinal density profiles of the gas jet along the laser axis for two durations of pre-pulse irradiation preceding the main laser pulse (the unperturbed density profile is shown in black). The blue profile corresponds to conditions explored in the present experiment. The red profile corresponds to longer prepulse irradiation that would lead to a shorter density profile. (b) Plasma profiles at t=5.7 ps (0.8 ps before the peak of the main pulse reaches x=0 mm). Both profiles are obtained for an irradiation by a laser pulse having an intensity characterized by $a_0$=4.*

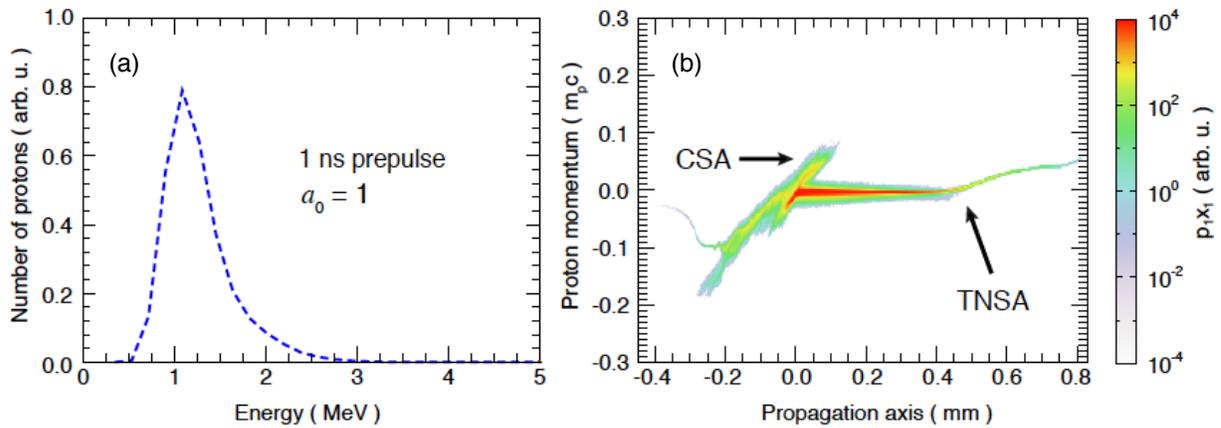

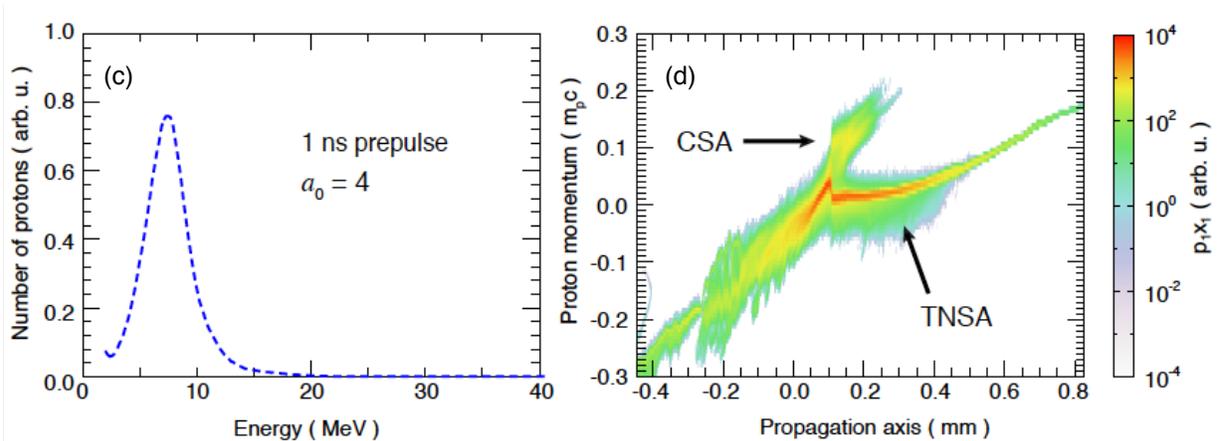

*Figure 7: Simulated proton properties obtained with the blue plasma profile shown in Fig.6 at t=11.7 ps. (a) Spectrum of the shock accelerated protons using a laser intensity of $a_0$=1. (b) Corresponding proton longitudinal phase space with a laser intensity of $a_0$=1. (c) Spectrum of the shock accelerated protons using a laser intensity of $a_0$=4. (d) Corresponding proton longitudinal phase space with a laser intensity of $a_0$=4.*

We have tested in the simulation irradiating the gas jet at two laser intensities, namely $a_0$=1 and $a_0$=4. We chose these two intensities as they correspond to the maximum intensity case ($a_0$=4, i.e. in the plane of best focus of the laser), or the case of reduced intensity ($a_0$=1, resulting from the laser defocus by 150 μm due to the gas jet deformation induced by the prepulse). In both cases, a clear shock structure has formed at the target critical density interface irradiated by the laser. The resulting phase space of the accelerated protons for our experimental conditions is illustrated for the two laser intensities in Fig.7b and Fig.7d. We observe that for both laser intensities, the phase space exhibits TNSA accelerated protons, those from hole-boring, as well as CSA accelerated ones which corroborates the fact that we observe several energy distributions in the experimental spectrum. The higher energy CSA accelerated protons lead to, as shown in Fig. 7a and Fig.7c, peaks in the spectrum. Here, the TNSA accelerated proton spectrum is not included to highlight the population accelerated by CSA, and its correspondence to the experimentally observed spectral peak corresponding to a peak density of 2.5 $n_{cr}$, which is shown (red curve) in Fig.4a. In the first case the peak is close to 1 MeV, i.e. consistent with the red curve measurement shown in Fig.4a, which supports our conjecture of the laser beam being indeed defocused to such a reduced intensity. This is further supported by the fact that the energy of the protons in the simulation using $a_0$=4 is higher than the one recorded in the experiment.

Figure 8 shows the temporal evolution of the electron density and allows us to read directly the velocity of the density discontinuities in units of *c*, the speed of light. This is done for the two cases of the two density profiles of the gas jet shown in Fig.6 of the paper as modified by the ASE of the Titan short-pulse laser (having $a_0$=1 in both cases). The interaction with the laser prompts a partial expansion backwards of the initially underdense sections of the plasma profiles. One observes that one cannot therefore clearly define a single acceleration velocity from Fig. 8, because of the density gradients in the system. However, there are several density discontinuities that propagate into the gas jet. The white line highlights the dominant one (corresponding ion phase spaces are shown in Figs 7b and 9a). The velocity of the maximum density peak in Fig. 8a is $v = 0.017c$. If we assume the ions reflected from this peak would have the velocity $v_i = 2v$, the ion energy is ~ 0.54 MeV. The velocity in Fig. 8b is somewhat higher, $v = 0.024c$, so the reflected ions are expected to be around 1 MeV. We note that these values are consistent with the spectrum of the reflected ions shown in Fig. 7a and 9c and which correspond to the conditions in which these simulations were run (i.e. to a gas density of 2.5 $n_{cr}$).

We also note that the proton energy of ~ 0.54 MeV that would result from the density discontinuity motion seen in Fig.8a is in reasonable agreement with the red spectrum shown in Fig.4a, which is recorded in experimental conditions corresponding to the ones of the simulation, i.e. using a peak density of the gas jet of 2.5$n_{cr}$, and at $a_0$~1. Such reasonable correspondence between the simulated and measured proton energy suggests that reflection of ions on the discontinuity, i.e. the CSA mechanism, observed in the simulations is indeed at play in the experiment.

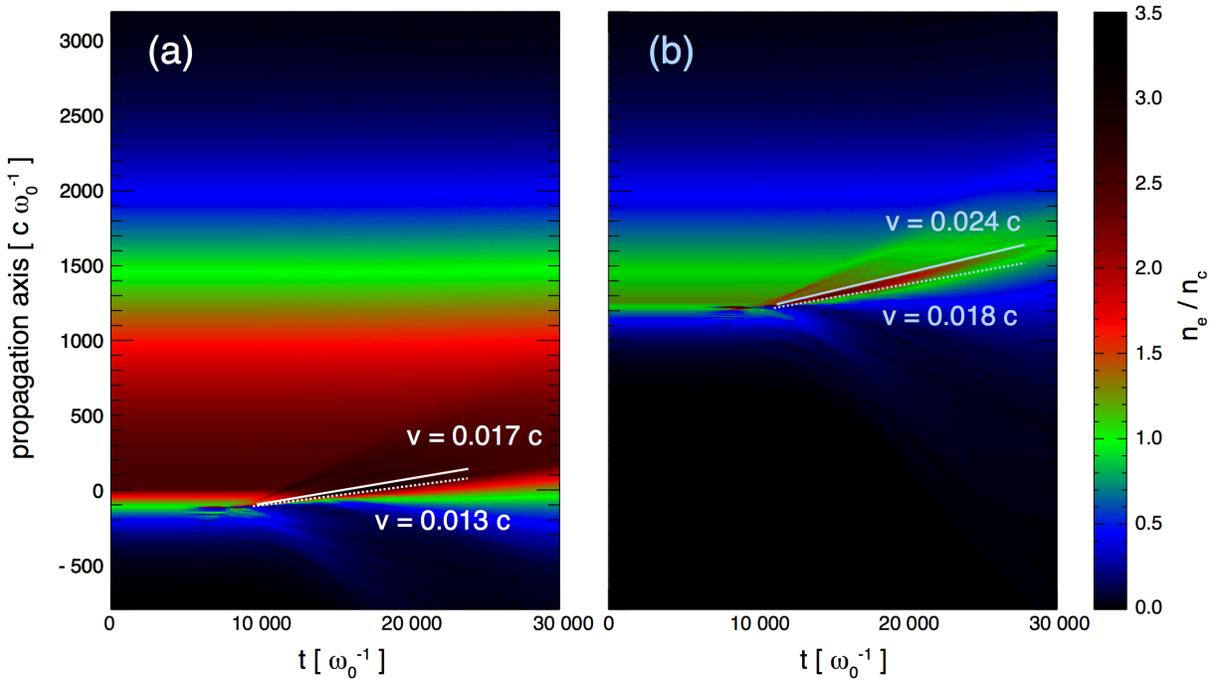

*Figure 8: Electron density evolution as a function of time extracted from the PIC simulations. Here, the time is normalised to the laser frequency $\omega_0$, and the propagation length to the $c/\omega_0$, where c is the speed of light. The initial density profiles correspond, respectively, to the (a) blue, and (b) red profiles shown in Fig. 6. The laser normalized intensity is $a_0=1$ in both cases. The white dashed lines now show the hole-boring velocity, and full lines the corresponding shock velocity.*

Moreover, we could analyse in the simulations at the velocity of the recessing point at which laser field reflection occurs, i.e. the HB velocity. It yields the following for the simulations discussed above.

|  | $v_{hb}/c$ | $v_{sh}/c$ |
|---|---|---|
| $a_0 = 1$, and using the blue profile of Fig.6 (1 ns ASE)and | 0.013 | 0.017 |
| $a_0 = 1$, and using the red profile of Fig.6 (1.9and ns ASE) | 0.018 | 0.024 |
| $a_0 = 4$, and using the blue profile of Fig.6 (1 ns ASE) | 0.049 | 0.065 |
| $a_0 = 4$, and using the red profile of Fig.6 (1.9 ns ASE) | 0.08 | ≥ 0.1 c (here we have several density discontinuities in the system with different velocities) |

Table 1: Measured velocities of hole-boring (HB) and of the electron discontinuity observed to propagate in the simulations (shock, SH, as illustrated in Fig.8), for various conditions, as stated.

Hence, the simulations clearly demonstrate that, in conditions of high density (all these simulations are performed with 2.7 $n_{cr}$ as the peak density), the dominant electron populations are characterized by a density spike propagating faster than the hole-boring (and in front of it). This is another indication that indeed CSA is at play here.

Finally, we have gauged these velocities with respect to the measured plasma electron temperature in the simulations to verify that it did not affect shock formation, or the shock velocity. The plasma electron temperature ($T_e$) is here measured in the upstream region at the time when the hole boring starts and the shock is formed. For $a_0=1$, $T_e \sim 0.12$ MeV (for both the red and blue profiles of Fig.6), yielding a sound speed around $c_s=0.011$ c. Stockem-Novo et al. [73] state that for near-critical density targets, the shock is launched if $v_{hb} > c_s$, which is the case here, referring to the $v_{hb}$ given in Table 1. Moreover, as $v_{hb} \ll c$, we expect that $v_{sh} / v_{hb} = 4 / 3$, which is indeed verified in Table 1. For $a_0=4$, $T_e \sim 1$ MeV (at maximum, i.e. at the peak of the laser irradiation on target), yielding $c_s = 0.033$ c. Again, we verify (see Table 1) $v_{hb} > c_s$, and that as well $v_{sh} / v_{hb} \sim 4 / 3$. This further corroborates that shock acceleration is here at play, with velocities following theoretical scalings.

**Conclusions**

We have demonstrated the ability to accelerate protons through possibly the CSA process with a 1.054 μm laser and we have observed that there are certainly trends that should be emphasized since they greatly affect the efficiency of CSA. First, we should note that the features in the spectrum are controllable by changing the peak density in the gas jet, and can be optimized (see below) by reducing its thickness. Indeed, we had observed that as the density of the gas jet increased, so did the peak energy of the quasi-monoenergetic bunch. Furthermore, the minimum required density to observe a peaked proton beam was, in our case where we used a laser with wavelength of 1.054 μm, above $n_{cr} = 1x10^{21}$ $1/cm^3$. Thirdly, the angular distribution is also sensitive to the gas jet density; we observed that the higher the density, the broader the angular distribution.

Interestingly, the PIC simulations point out to, at high densities, a CSA acceleration mechanism since the highest energy protons are accelerated by a density spike that travels through the target at a velocity higher than the HB one. The experimental data at high density are seen also to be close to existing CSA analytical scalings. This last point could be of interest for assessing focal intensity on target at future high-intensity facilities (GIST, APOLLON, ELI) for which such measurement at the actual target location is still a challenge.

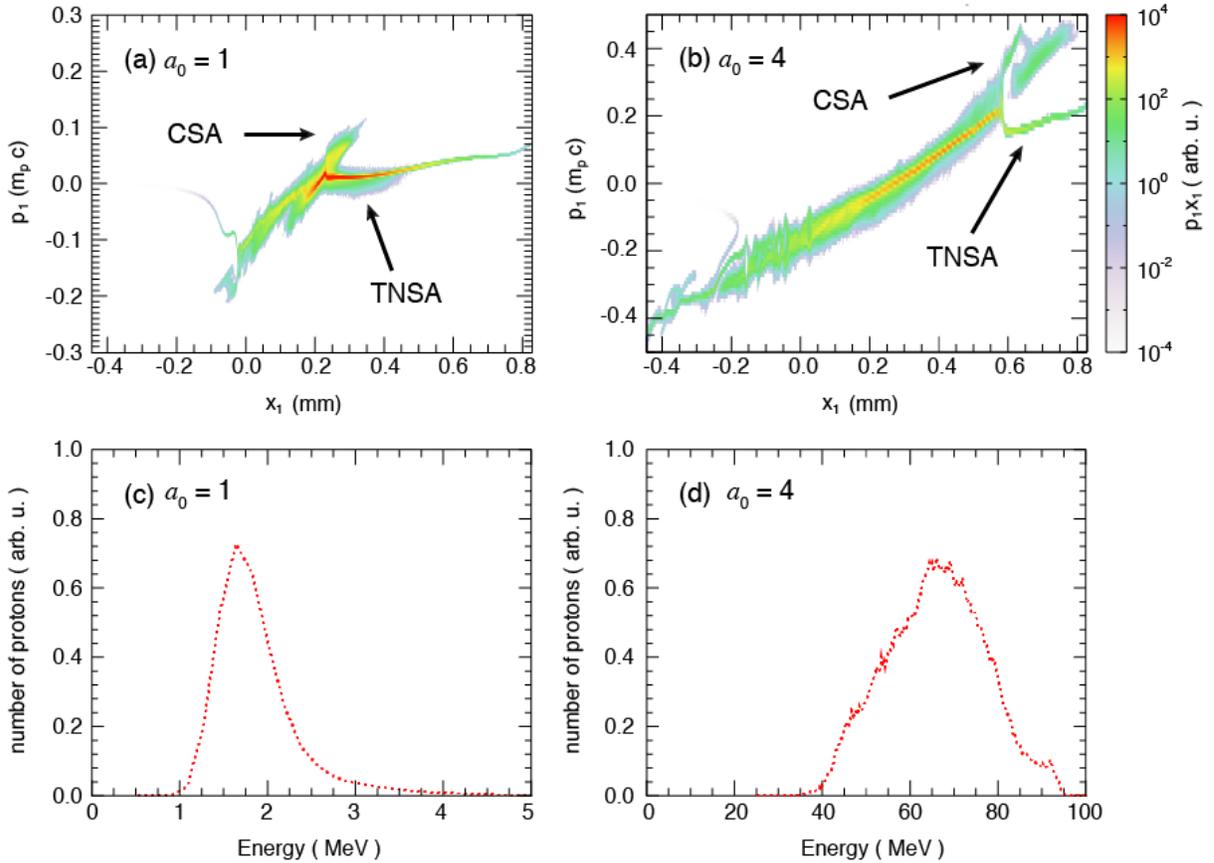

*Figure 9: Proton properties obtained, for two different laser intensities (as labeled), with the plasma profile corresponding to the red profile shown in Fig.6 (i.e. corresponding to a 1.9 ns prepulse) at t=11.7 ps. (a)-(b) Proton longitudinal phase space. (c)-(d) Spectrum of the shock accelerated protons.*

As a perspective, we have explored the effect of further reducing the thickness of the near-critical density profile in the target. For this, we tested (using the hydrodynamic code) what was the effect of having an even longer prepulse interacting with the gas jet. This is shown in Fig 6a: the red profile shows that when prolonging up to 1.9 ns the prepulse irradiation (i.e. longer than actually used in the experiment), we can indeed shorten even further the gas jet profile. Of course, this would induce the main laser pulse to be even more defocused. This was compensated in the simulations shown in Fig.8 by keeping the laser intensity on the critical density interface at $a_0=1$ or $a_0=4$. When using this shortened profile and these laser intensities, the resulting proton beam parameters obtained in the PIC simulations are shown in Fig.8. It can be seen that reducing the thickness of the plasma, i.e. the amount of plasma on the back side, can indeed improve significantly the final proton energy. In the case with a thicker target (as explored in the experiment and shown in Figs. 6 and 7), the proton energy is low, but the spectral bandwidth is small. In the case where the target is thinned out, with also a decrease of 2.5 times of the peak target density, we observe that we can obtain much higher final proton energy (see Fig.8d), but here with some cost on the bandwidth, which is significantly larger. Testing such reduced width critical density gas jet will be explored in further experiments, either by changing the prepulse parameters or by using directly thinner gas jets.

**Acknowledgements**

The authors thank the staff of the Titan Laser and the Jupiter Laser facility for their support during the experimental preparation and execution. We thank R. C. Cauble and J.R. Marquès for fruitful discussions, and R. Riquier for the gas jet characterization. This work was partly done within the LABEX Plas@Par project and supported by Grant No. 11-IDEX-0004-02 from Agence Nationale de la Recherche. It has received funding from the European Union's Horizon 2020 research and innovation programme under grant agreement no 654148 Laserlab-Europe, and was supported in part by the Ministry of Education and Science of the Russian Federation under Contract No. 14.Z50.31.0007. The use of the Jupiter Laser Facility was supported by the U.S. Department of Energy by Lawrence Livermore National Laboratory under Contract DE-AC52-07NA27344. This work is supported by FRQNT (nouveaux chercheurs, Grant No. 174726, Team Grant 2016-PR-189974), NSERC Discovery Grant (Grant No. 435416), ComputeCanada (Job: pve-323-ac, P. Antici). M.V. and L.O.S. were supported by the European Research Council (ERC-2010-AdG grant no. 267841). O.Willi acknowledges the DFG Programmes GRK 1203 and SFB/TR18. M. B.-G., J. J. S., and E. d'H. acknowledge the funding from the project ARIEL (Conseil Regional d'Aquitane); their work was carried out in the framework of the Investments for the Future Programme IdEn Bordeaux LAPHIA (ANR-10-iDEX-03-02). The PIC simulations were performed at Fermi/Marconi (Italy) through PRACE allocation and at the Accelerates cluster (Lisbon, Portugal).

**Author Contributions**

JF, SNC and EdH conceived the projet. SNC wrote the manuscript, made figures 1, 2, 4, & 5 organized and lead the experiment, and analysed the data from the proton spectrometers; MV performed the PIC simulations, as well as their analysis, and made figures 6, 7, 8 and 9; TG prepared the experiment, laser and diagnostic alignment and characterized the gas nozzles; EB gave valuable insight in the interpretation of the PIC simulation results and performed post-processing of the results; PA participated in the experiment; MBG fielded diagnostics during the experiment; PL performed the hydrodynamic simulations using FCI2 and made figure 3; HP supervised students and participated in the experiment; GR prepared the experiment, laser and diagnostic alignment; JJS supported the experiment and students financially; AMS participated in the experiment and fielded diagnostics; MS supervised the data analysis and revised the manuscript; OW supervised students and provided financial support, JF supervised fielding the experiment, the data analysis and helped write the manuscript. All authors reviewed the manuscript.

**Additional Information**

**Competing financial interests**

The Authors of this article declare no competing financial interests.

rugby hohlraum on the Omega Laser Facility: Comparisons between cylinder, rugby, and elliptical hohlraums. *Phys. Plasmas* **23**, 022703 (2016).
72 Willingale, L., et al., Collimated Multi-MeV Ion Beams from High-Intensity Laser Interactions with Underdense Plasma. *Phys. Rev. Lett*. **96**, 245002 (2006).
73 Stockem Novo, A., Kaluza, M. C., Fonseca, R. A., Silva, L.O., Optimizing laser-driven proton acceleration from overdense targets. *Sci. Rep*. **6**, 29402 (2016).
74 Ping, Y., et al., Absorption of Short Laser Pulses on Solid Targets in the Ultrarelativistic Regime. *Phys. Rev. Lett*. **100,** 085004 (2008).
75 Fonseca, R. A., *et al*., OSIRIS: A Three-Dimensional, Fully Relativistic Particle in Cell Code for Modeling Plasma Based Accelerators. *Lecture Notes in Computer Science* **2331**, 342 (2002).